\documentclass[%
 reprint,
 amsmath,amssymb,
 aps,
]{revtex4-2}

\usepackage{float}
\usepackage[normalem]{ulem}
\usepackage{graphicx}
\usepackage{dcolumn}
\usepackage{bm}
\usepackage{xcolor}
\usepackage{hyperref}

\begin{document}

\preprint{APS/123-QED}

\title{Programming tunable active dynamics in a self-propelled robot}

\author{Somnath Paramanick$^{1}$}
\author{Arnab Pal$^{2}$}
\author{Harsh Soni$^{3}$}
\author{Nitin Kumar$^{1}$}
\email{nkumar@iitb.ac.in}
\affiliation{
	$^1$Department of Physics, Indian Institute of Technology Bombay Powai, Mumbai 400076, India \\
 $^2$The Institute of Mathematical Sciences, CIT Campus, Taramani, Chennai 600113, India \& Homi Bhabha National Institute, Training School Complex, Anushakti Nagar, Mumbai 400094, India\\
	$^3$School of Physical Sciences, Indian Institute of Technology Mandi, Mandi 175001, India \\
	}
\date{\today}






\begin{abstract}
We present a scheme for producing tunable active dynamics in a self-propelled robotic device. The robot moves using the differential drive mechanism where two wheels can vary their instantaneous velocities independently. These velocities are calculated by equating robot's equations of motion in two dimensions with well-established active particle models and encoded into the robot's microcontroller. We demonstrate that the robot can depict active Brownian, run and tumble, and Brownian dynamics with a wide range of parameters. The resulting motion analyzed using particle tracking shows excellent agreement with the theoretically predicted trajectories. Finally, we demonstrate that its motion can be switched between different dynamics using light intensity as an external parameter. This work opens an avenue for designing tunable active systems with the potential of revealing the physics of active matter and its application for bio- and nature-inspired robotics.

\end{abstract}

\maketitle


\section{INTRODUCTION}
Active matter refers to systems comprised of self-propelled particles that consume energy to perform mechanical work \cite{ramaswamy2010mechanics,marchetti2013hydrodynamics,chate2020dry,toner2005hydrodynamics}. Such systems exhibit striking non-equilibrium phenomena such as collective dynamics, self-organization, and anomalous mechanical properties ubiquitous in the natural and biological world \cite{kumar2014flocking,sanchez2012spontaneous,brugues2014physical,cates2015motility,kumar2022catapulting,palacci2013living,brugues2014physical,kumar2018tunable,simha2002hydrodynamic,hatwalne2004rheology,rafai2010effective,lopez2015turning,fruchart2023odd,nash2015topological}. Over the past few decades, numerous analytical, numerical, and experimental studies have advanced our understanding of such systems. Yet we are just beginning to realize the potential of active matter in building smart, adaptable materials and autonomous devices \cite{brandenbourger2019non,souslov2017topological,gossweiler2015mechanochemically}. Achieving this will require better control over individual particle dynamics and inter-particle interactions. Thus, many studies in recent years have reported concerted efforts to achieve programmable control on active matter spanning a broad range of length scales \cite{unruh2023programming,PatterenedActivity2021,buttinoni2012active,li2019particle,fernandez2020feedback,prasath2022dynamics,siebers2023exploiting,nakayama2023tunable,unruh2023programming,aubert2017evolutionary,zeravcic2017colloquium,zhang2021spatiotemporal,lemma2022spatiotemporal}. Here, we develop a novel active matter system consisting of a collection of smart programmable self-propelled robots capable of sensing their environment using onboard sensors. As a first step, in this paper, we present simple protocols for extracting tunable active dynamics in a single robot.

Previous studies have shown that emerging properties in active systems depend on the self-propulsion mechanism of individual agents, the strength of alignment interaction, and the nature of the surrounding medium \cite{marchetti2013hydrodynamics}. Therefore, an artificial active material must incorporate all these features to replicate complex non-equilibrium properties displayed by active systems. Yet, broadly speaking, modeling individual constituent dynamics using simple theoretical frameworks of active Brownian particles (ABP) and run-and-tumble particles (RTP) has proven successful in replicating emerging phenomena with great success \cite{chate2020dry}. Thus, a robotic device, due to its programmability, becomes a natural choice to mimic controlled active particle dynamics. While there are many examples in the literature where robots perform programmable complex self-assembly \cite{rubenstein2014programmable} as well as spontaneous collective behavior \cite{deblais2018boundaries,yang2020robust,dmitriev2021statistical,giomi2013swarming} they rely on vibration motors for locomotion \cite{rubenstein2014programmable,deblais2018boundaries,yang2020robust}. More recently, differential drive robots have also been reported which show much better control of individual dynamics \cite{wang2021emergent,leyman2018tuning,arvin2017mona,prasath2022dynamics}. Yet, a detailed protocol to program them with well-established theoretical models of ABP, RTP, and Brownian particle (BP) is missing. 

\begin{figure*}[t]
	\centerline{\includegraphics[width=0.6\textwidth]{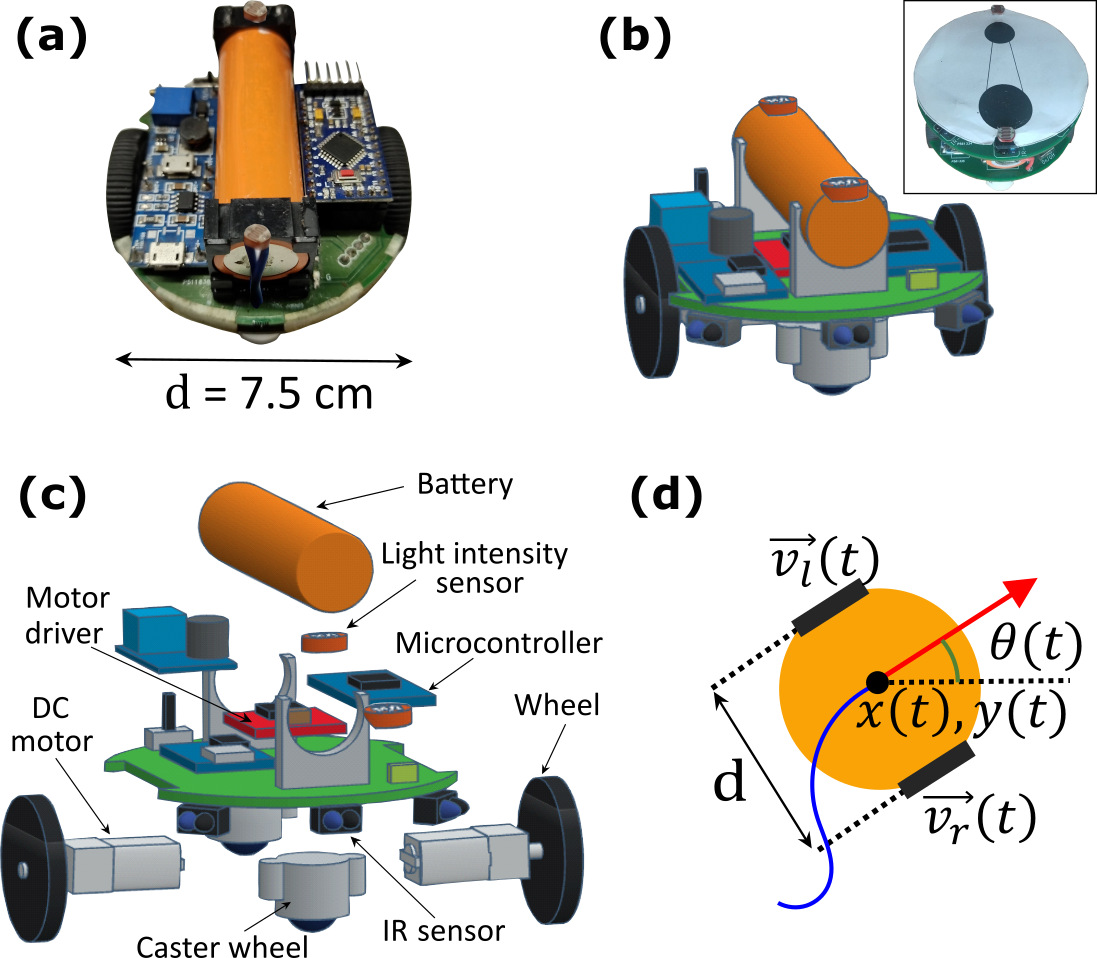}}
	\caption{\textbf{Robot model}: (\textbf{a}) The robot used in the experiment with its diameter indicated. (\textbf{b}) A schematic representation of the same robot. The inset shows the top view of the robot covered with a white cap with two black circles on top along its orientation. (\textbf{c}) An expanded view highlighting its important hardware components. (\textbf{d)}  Robot's position $\textbf{(}x(t),y(t)\textbf{)}$ and orientation $\theta(t)$ vary as a function of the instantaneous left and right wheel's velocities ($v_l(t)$ and $v_r(t)$, respectively), resulting in the centre of mass tracing out a trajectory indicated by the blue solid line.}
	\label{fig1}
\end{figure*}

In this work, we present a robotic system that displays a variety of active dynamics with many controllable parameters. Our robot is circularly shaped and propels itself using the differential drive mechanism. In addition, the robot also houses hardware components like multiple infrared (IR) and light intensity (LI) sensors which help it detect environmental factors like obstacles and ambient light. We demonstrate that with its two wheels oppositely placed at a diameter distance away, its center of mass can display a variety of motions depending entirely on the instantaneous wheel velocities. In particular, we show our results on the following three models: (i) Active Brownian particle (ABP) with and without translational noise, (ii) Run and Tumble particle (RTP), and (iii) Brownian particle (BP) dynamics. These results are found to be in excellent agreement with our theoretical modeling. We also show that the robot can be programmed to switch between these dynamics as a function of the ambient light intensity, providing spatiotemporal control over its motion. Taken together, our results propose a novel experimental system based on smart programmable robots that provide an ideal platform for exploring problems related to active and living systems with potential in future material design.

\section{ROBOT MODEL}
Our robot is a circularly shaped electronic gadget with a diameter of 7.5 cm and a height of 5 cm (See Fig.~\ref{fig1}(a-c)). It is assembled using various hardware components as indicated in Fig.~\ref{fig1}(c). It has two rubber wheels of diameter 3.3 cm placed at the two ends of its diameter. Two caster wheels are placed in the perpendicular direction for support. It also carries six IR sensors symmetrically placed along the circumference to detect obstacles. Two light intensity (LI) sensors, each placed at the front and back, enable the robot to detect the ambient light intensity gradients. We use the differential drive technique for the robot's movement on a 2D plane surface by controlling the velocity of the wheels independently \cite{malu2014kinematics}. The programs, written using Arduino IDE software, are loaded onto an Arduino pro mini microcontroller board (3.3 V, 8 MHz version), and the output signal is fed to the motor driver (Dual TB6612FNG by SparkFun), that rotates the wheels using two separate DC motors. The signal from the microcontroller provides 256 different analog voltage levels from zero to maximum voltage (5V) to the motors. This gives each wheel a range of velocities maximum of up to 27 cm/sec. All these components are powered using a Li-ion battery. A brief summary of robot functioning is presented in SI Fig.1. The robot performs its dynamics on a flat surface of dimensions 90 cm $\times$ 120 cm with a white background. We use an overhead projector connected to a computer to illuminate the surface with the desired light intensity. We capture the dynamics of the robot using a high-speed camera from the top and perform image analysis to extract instantaneous position and orientation (See Appendix A).\\

We begin by writing the equations of motion of the robot in two dimensions. Let $v_l(t)$ and $v_r(t)$ be the velocities of the left and right wheels of the robot along its orientation at time $t$ (See Fig.~\ref{fig1}(d)). Then, the equations of motion for the position $\textbf{(}x(t),y(t)\textbf{)}$ and the orientation angle $\theta(t)$ of the robot read 
\begin{equation}\label{eqn1}
\begin{aligned}
\dot x (t)&= \frac{v_l(t) + v_r(t)}{2} \cos \theta(t),\\
\dot y(t)&= \frac{v_l(t) + v_r(t)}{2}\sin \theta(t),\\
\dot \theta(t)&= \frac{v_r(t) - v_l(t)}{d},
\end{aligned}
\end{equation}
where $d$ is the diameter of the robot, and the dot stands for the time derivative. We next show that purely based on the instantaneous values of $v_l$ and $v_r$, we can program the robot to perform desired stochastic dynamics. $\frac{1}{24}$
\begin{figure*}
	\centerline{\includegraphics[width=0.9\textwidth]{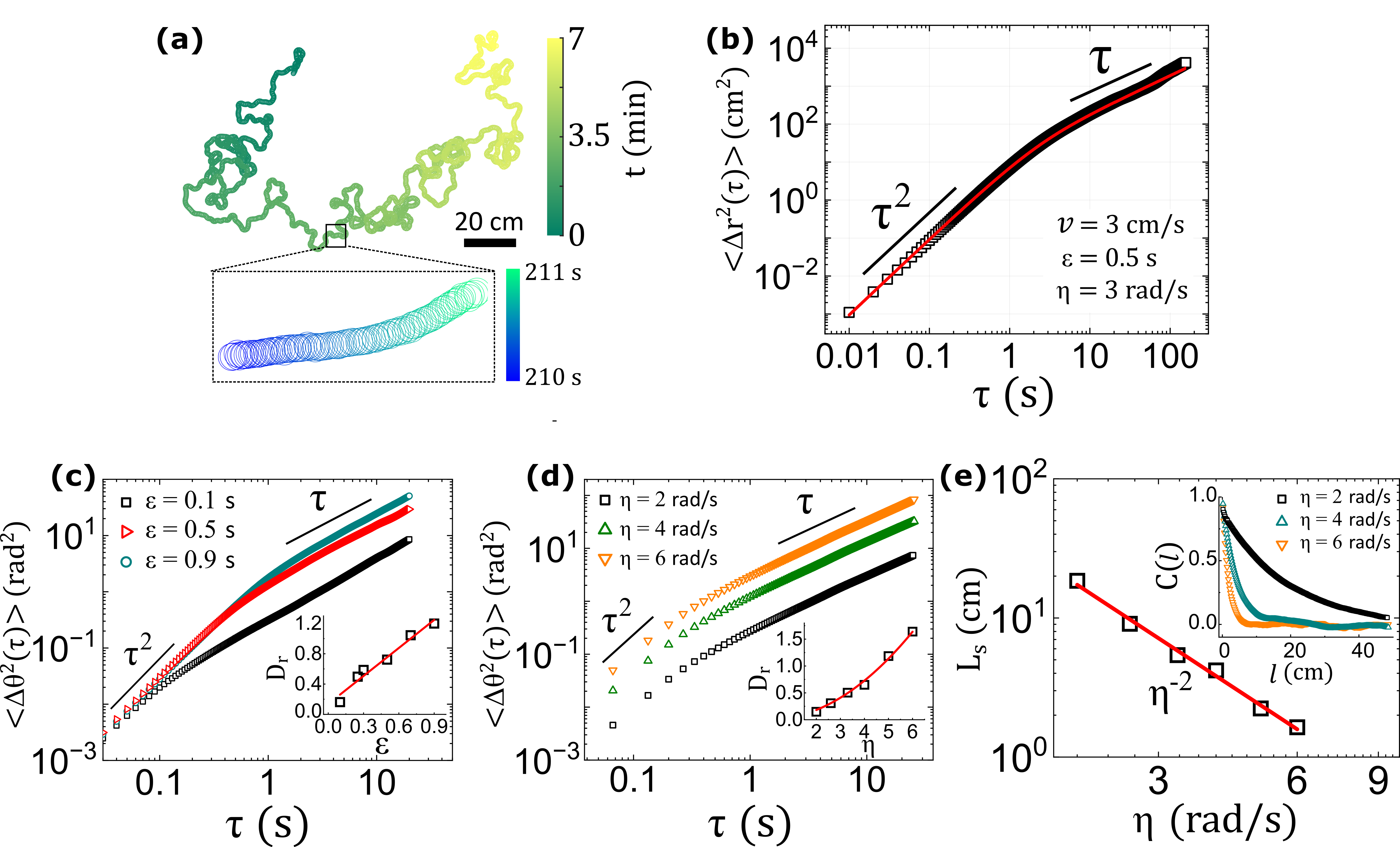}}
	\caption{\textbf{ABP model without translational noise}: (\textbf{a}) A typical trajectory of the robot following ABP dynamics with velocity $v$ = 3 cm/s, delay time $\epsilon$ = 0.5 s, and rotational noise $\eta$ = 3 rad/s. The color bar represents the time. The zoomed-in portion of the trajectory shows the robot moving with a constant velocity at all time steps, indicative of the absence of translational noise at short time scales. (\textbf{b}) Translational mean-square displacement (MSD) of the trajectory plotted as a function of time $\tau$ showing a ballistic ($\sim\tau^2$) to diffusive ($\sim\tau$) transition. The solid line shows the fit with the analytical expression given by~\eqref{eqn4}. (\textbf{c}) Angular MSD, $< \Delta\theta ^2(\tau)>$, plotted for three values of $\epsilon$ keeping $\eta$ = 3 rad/s. $< \Delta\theta ^2(\tau)>$ also undergoes a ballistic to diffusive transition at time $\epsilon$. (Inset) The experimental values of $D_r$, measured by line fitting the linear part of the MSD, show linear scaling with $\epsilon$. (\textbf{d}) Angular MSD as a function of $\tau$ for different values of $\eta$ keeping $\epsilon$ fixed at 0.25 s. (Inset) $D_r$ extracted from the plots increases quadratically with $\eta$. (\textbf{e}) Persistence length $L_s$ shows an algebraic decay with $\eta$ with an exponent consistent with theory. $L_s$ is extracted by fitting an exponential function to the spatial velocity autocorrelation function, $C(l) = \left\langle\mathbf{v}(r).\mathbf{v}(r+l)\right\rangle_r / \mathbf{v^2}$ shown in the inset.}
	\label{fig2}
\end{figure*}
\section{RESULTS AND DISCUSSION}
\subsection{ABP model without translational noise}\label{ABPM}
A distinctive feature of a polar active particle is its finite self-propulsion speed. In general, it can have rotational as well as translational noise. We first study the case without an explicit source of translational noise. In that case, it is often modeled as an ABP, which is equivalent to a biased random walk in two dimensions where the particle always moves in the direction of its instantaneous orientation. Consequently, the ABP moves with a constant speed but its direction changes randomly in the plane. The dynamics of an ABP of speed $v$ is governed by the following equations of motion:
\begin{equation}\label{eqn2}
\begin{aligned}
\dot x(t)&= v \cos \theta(t),\\
\dot y(t)&= v \sin \theta(t), \\
\dot \theta(t)&= \zeta(t),
\end{aligned}
\end{equation}
where $\zeta$ is the instantaneous rotational speed of the ABP; it is a white noise with a zero mean and is often modeled as white Gaussian noise. Here, we construct an active Brownian robot with $\zeta$ having a uniform distribution between $ [-\eta, \eta]$. In order to produce an ABP motion of given $\eta$ and $v$ in our robot, we compare~\eqref{eqn2} and~\eqref{eqn1} to obtain the expressions of $v_l$ and $v_r$, which yields
\begin{equation}\label{eqn3}
\begin{aligned}
v_l(t)&= \frac{2v + \zeta(t) d}{2},\\
v_r(t)&= \frac{2v - \zeta(t) d}{2}.\\
\end{aligned}
\end{equation}
The values of $v_l$ and $v_r$ are updated in the robot after every $\epsilon$ seconds, termed as delay time. In experiments, the accessible values of parameters $\eta$ and $\epsilon$ lie between 1 - 6 rad/s and 0.1 - 0.9 s, respectively.

When implemented in the robot, it mimics the ABP motion quite well. A typical trajectory of the robot is shown in Fig.~\ref{fig2}(a) and SI movie S1. To get this unbounded free particle trajectory, we use boundary wall detection using IR sensors to remove the confinement effects (See Appendix B). The zoomed-in part of the trajectory ascertains the absence of translational noise at short time scales. The mean squared displacement (MSD) shows a transition from ballistic to diffusive behavior, a hallmark feature of the ABP dynamics (see Fig.~\ref{fig2}(b)). Using a simple theoretical calculation, we obtain the following expression for the MSD~\cite{howse2007self,bechinger2016active}
\begin{equation}\label{eqn4}
\begin{aligned}
\left\langle\Delta r^2(\tau)\right\rangle = \frac{2v^2}{D_r}[\tau + \frac{1}{D_r}(e^{-D_r \tau} - 1)],
\end{aligned}
\end{equation}
 where $D_r = \epsilon \left\langle\zeta^2\right\rangle /2= \epsilon \eta^2/6$ (see Appendix C). The solid red line in Fig.~\ref{fig2}(b) represents analytically obtained MSD, which shows an excellent agreement with our experimental data. We further plot angular MSD defined as $\left\langle \Delta\theta ^2(\tau)\right\rangle$ for varying $\epsilon$ and $\eta$ independently (See Figs.~\ref{fig2}(c) and ~\ref{fig2}(d)). Interestingly, we find that it scales linearly with $\tau$ only for $\tau > \epsilon$, whereas for $\tau < \epsilon$, the plot is ballistic with $\tau^2$ scaling. This happens because during the time duration $\epsilon$, the robot always rotates with a constant angular velocity of magnitude $\lvert(v_l-v_r)/d\rvert$, resulting in $\Delta\theta$ scaling linearly with $\tau$. For given $\epsilon$ and $\eta$, the experimental value of $D_r$ is measured by line-fitting $\left\langle \Delta\theta ^2(\tau)\right\rangle$, using the formula $\left\langle\Delta\theta^2(\tau)\right\rangle$ = $2D_r\tau$.  We then plot $D_r$ as a function of $\epsilon$ and $\eta$ in the insets of Figs.~\ref{fig2}(c) and ~\ref{fig2}(d), respectively, showing $D_r$ indeed scaling according to the expression $D_r = \epsilon \eta^2/6$, as predicted by the theory.
 
We also measure spatial velocity autocorrelation function, $C(l) = \left\langle\mathbf{v}(r).\mathbf{v}(r+l)\right\rangle_r / \mathbf{v^2}$,  where $r$ is the distance traveled by the robot, with respect to $l$ for different values of $\eta$ keeping $\epsilon$ constant at 0.25 s as shown in Fig.~\ref{fig2}(e) inset. We find that $C(l)\propto e^{-l/L_s}$, where $L_s$ is the persistence length. Using the above relation, we extracted $L_s$ for different $\eta$. We find that $L_s$ decreases algebraically with $\eta$ with an exponent of ``-2''. This also agrees with the theory since $L_s=v/D_r\propto \eta^{-2}$.
\subsection{ABP model with translational noise}
In general, at small time scales, a polar particle may have translational noise along as well as normal to its orientation. In this case, therefore, $v$ in~\eqref{eqn2} is not a constant and rather fluctuates around its mean value. We achieve this by generating $v_l(t)$ and $v_r(t)$ from two statistically independent random uniform distributions in a range of $2V$ with a finite mean $v_a$, i.e., in the interval $[v_a-V,v_a+V]$. Here $V$ quantifies the strength of the translational noise. Interestingly, since our robot, at any instance, can only move along the direction of its wheels, this noise features only in the velocity component, which is parallel to its orientation, while noise normal to its orientation is forbidden. As a result, both $\dot\theta(t)$ and $v(t)$ become random variables with zero and a non-zero mean (= $v_a$), respectively.

\begin{figure*}
	\centerline{\includegraphics[width=0.6\textwidth]{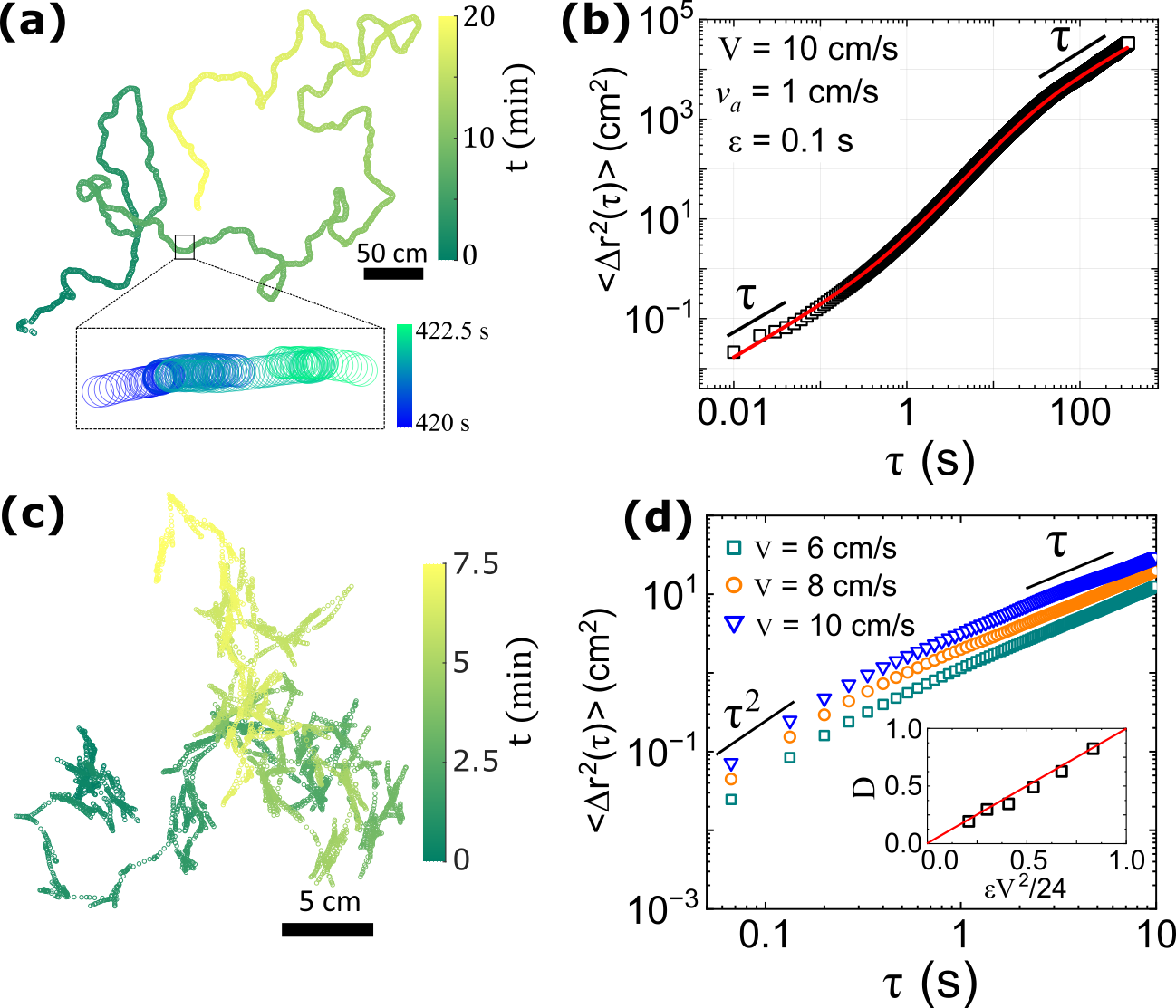}}
	\caption{\textbf{ABP model with translational noise and BP model}: (\textbf{a}) Typical trajectory of the robot programmed to follow ABP model with translational noise having $V$ = 10 cm/s, $\epsilon$ = 0.1 s, and $v_a$ = 1 cm/s. The color bar represents the time. The zoomed-in trajectory shows the presence of translational noise in the form of significant back-and-forth events along the robot's orientation. (\textbf{b}) The translational MSD shows a diffusive to super-diffusive to diffusive transition as a function of time. The solid line is fit with the analytical expression given by~\eqref{eqn5}. (\textbf{c}) A typical trajectory of the robot following a BP motion with $V$ = 7 cm/s, $\epsilon$ = 0.2 s, and $v_a$ = 0 cm/s. The color bar represents the time of motion. (\textbf{d}) The translational MSD of the BP motion plotted for three values of $V$ keeping $\epsilon$ fixed at 0.2 s shows a ballistic to diffusive transition. The ballistic motion is a consequence of the finite delay time $\epsilon$ which also sets the crossover time for all the curves. The inset shows the diffusion constant $D$ measured from the MSD curves matches with $\epsilon V^2/24$ as predicted by the theory.}
	\label{fig3}
\end{figure*}

Fig.~\ref{fig3}(a) and SI movie S2 show a typical trajectory of the robot following noisy ABP dynamics. At long time scales, the trajectory qualitatively looks similar to the ABP without translational noise (Fig.~\ref{fig2}(b)) but differs significantly at short time scales as depicted by frequent back-and-forth events in the zoomed-in trajectory. We plot experimentally measured MSD in Fig.~\ref{fig3}(b), which shows scaling with $\tau$ at both short and very long time scales as reported in past studies \cite{bechinger2016active}. We solve the Langevin equation for this case and obtain the following expression of the MSD
\begin{equation}\label{eqn5}
\begin{aligned}
\left\langle \Delta r^2(\tau) \right \rangle = 4D_t \tau + \frac{2v_a^2}{D_r^2}[D_r \tau + (e^{-D_r \tau} - 1)],
\end{aligned}
\end{equation}
where $D_t = \epsilon V^2/24$ and $D_r =  \epsilon V^2/3d^2$ are the translational  and the rotational diffusion constants, respectively (see Appendix C). Again the experimental data agree well with the analytical MSD expression (solid red line) over the entire range of time scales (Fig.~\ref{fig3}(b)). Clearly, in the limits $\tau \rightarrow 0$ and $\tau \rightarrow \infty$,~\eqref{eqn5} reduces to $\left \langle \Delta r^2(\tau)\right \rangle_{\tau \rightarrow 0} = 4D_t \tau$  and $\left \langle \Delta r^2(\tau) \right \rangle_{\tau \rightarrow \infty} = 4[D_t + v_a^2/(2D_r)] \tau$, respectively, showing an agreement with our experimental results. 

\subsection{Brownian particle model}
A Brownian particle (BP) in two dimensions can move in any direction with an equal probability. However, quite obviously, the wheels of our robot can move only along its direction of orientation, not normal to it. Here we show that it is possible for our robot to perform BP dynamics with limited movement of its wheels. Notably, the ABP motion with noise described in the previous section reduces to the BP dynamics when $v_a$ = 0. As a result, both $v(t)$ and $\dot\theta$ become random variables with zero mean. Fig.~\ref{fig3}(c) and SI movie S3 show the trajectory of the robot following the BP model with $V$ = 7 cm/s and $\epsilon$ = 0.2 s. The resulting MSD shows that at short times ($<\epsilon$) the motion is ballistic, and at longer times it is diffusive (see Fig.~\ref{fig3}(d)), similar to $\left\langle \Delta\theta ^2(\tau)\right\rangle$ observed in Figs.~\ref{fig2}(c) and ~\ref{fig2}(d). Again, the ballistic regime is a consequence of finite velocity events experienced by both left and right wheels over a time interval of $\epsilon$. This is further verified by the fact that the crossover from ballistic to diffusive regime happens exactly after $\epsilon$.

Moreover, since $\left\langle \Delta r ^2(\tau)\right\rangle = 4D\tau$, we extract the diffusion constants $D$ from the MSD plots for various values of $V$ and $\epsilon$. We plot this value with the analytical result, which predicts $D = \epsilon \left\langle v^2\right\rangle/4 = \epsilon  V^2/24$ in Fig.~\ref{fig3}(d) inset. Clearly, we find an excellent agreement with the theory (solid line indicating  the equality).

\begin{figure*}
	\centerline{\includegraphics[width=0.6\textwidth]{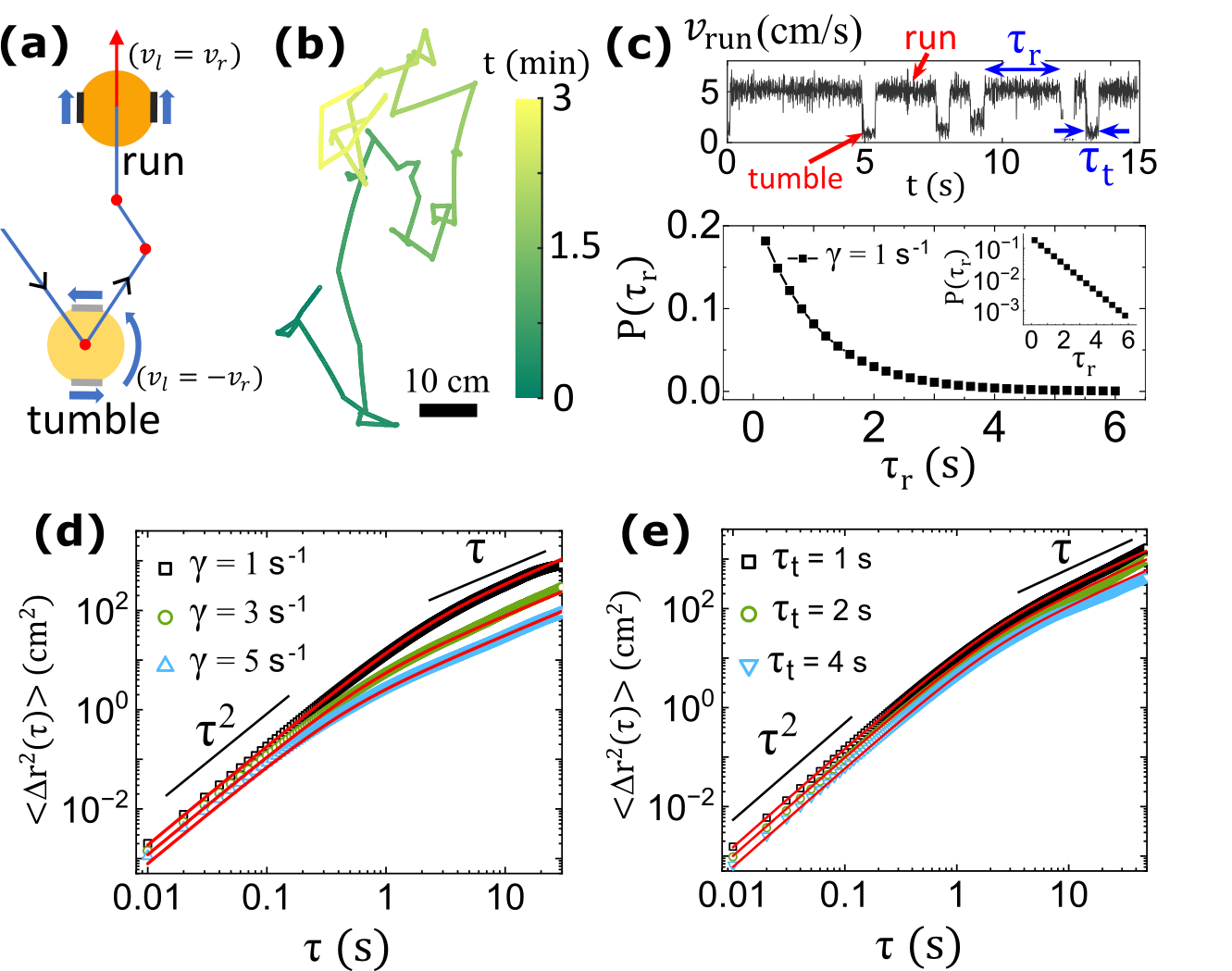}}
	\caption{\textbf{RTP model}: \textbf{(a)} A typical schematic of the robot mimicking RTP model. A run event is characterized by a straight trajectory when $v_l = v_r$. The robot tumbles while $v_l = -v_r$ and finds a new run direction randomly. \textbf{(b)} The trajectory of the robot for a duration of 3  minutes, programmed to follow RTP model. The color bar shows the time. \textbf{(c)} The top plot shows the instantaneous speed of the robot, which switches between $v_\text{run}$ = 5 cm/s and $v_\text{run}$ = 0, representing run and tumble, respectively. $\tau_r$ and $\tau_t$ represent run and tumble duration, respectively. $\tau_t$ is constant during the motion, whereas $\tau_r$ is extracted from an exponential distribution $P(\tau_r) = \gamma e^{-\gamma \tau_r}$ as shown in the bottom plot. A linear semi-log plot in the inset verifies the exponential distribution. Parameter values for the trajectory shown in \textbf{(b)} are $v_\text{run}$ = 5 cm/s, $\gamma$ = 3 s$^{-1}$, and $\tau_t$ = 0.5 s. \textbf{(d)} Translational MSD as a function of time for three different $\gamma$ values keeping $\tau_t$ = 0.5 s. Solid pink lines are the fit with the theoretical expression given by~\eqref{eqn6}. \textbf{(e)} Translational MSD for three different $\tau_t$ values keeping the $\tau_r$ distribution fixed with $\gamma$ = 1 s$^{-1}$. Again, solid lines represent the fit with the same theoretical expression.}
	\label{fig4}
\end{figure*}
\subsection{Run and tumble particle model}

We now turn our attention to the RTP models, where an individual particle moves in a straight line interrupted by frequent tumble events that randomize its run direction (Fig.~\ref{fig4}(a)) \cite{tailleur2008statistical,solon2015active}. Such motion is widely observed in micro-organisms like bacteria \cite{berg1972chemotaxis} and even in active granular particles \cite{kumar2019trapping}. In the case of our robot, it runs when the left and right wheels have the same velocities, i.e., $v_l=v_r = v_\text{run}$, and it tumbles to a new direction when $v_l=- v_r$. A typical trajectory of the robot performing RTP motion is shown in Fig.~\ref{fig4}(b) and SI movie S4 with its instantaneous speed plotted in Fig.~\ref{fig4}(c). Here, the robot runs with a constant speed $v_\text{run}$ for a duration $\tau_r$ (run time), after which it tumbles over a time $\tau_t$ (tumble duration). It is important to highlight that $\tau_t$ can never be zero in our experiments since the robot can never tumble instantaneously. Inspired by the experimental observations in bacteria \cite{berg1972chemotaxis}, we choose $\tau_r$ from an exponential distribution $P(\tau_r) = \gamma e^{-\gamma \tau_r}$, whereas $\tau_t$ is kept constant for all tumble events. This is achieved by scaling the tumble speed with the tumble angle, which is chosen randomly between 0 to 360 degrees. The RTP trajectory shown in Fig.~\ref{fig4}(b) has $v_\text{run}=5$ cm/s, $\gamma=3$ s$^{-1}$, and $\tau_t=0.5$ s. Later, we perform experiments for various $\tau_r$ distributions characterized by $\gamma$ at fixed $\tau_t$ = 0.5 s. The resulting MSD is shown in Fig.~\ref{fig4}(d), showing a ballistic ($\sim \tau^2$) to diffusive ($\sim \tau$) transition with transition time decreasing as we increase $\gamma$. However, when we vary $\tau_t$ while keeping $\gamma$ constant (Fig.~\ref{fig4}(e) for $\gamma$ = 1 $\text{s}^{-1}$), the transition time remains independent of $\tau_t$. Regardless, the MSD measured experimentally shows an excellent agreement with the theoretical expression given by \cite{angelani2013averaged}
\begin{equation}\label{eqn6}
\begin{aligned}
\left \langle \Delta r^2(\tau) \right \rangle = \frac{1}{1 + \gamma \tau_t}\frac{2v_\text{run}^2}{\gamma^2}(\gamma \tau - 1 + e^{-\gamma \tau}).
\end{aligned}
\end{equation}
The solid red lines in Fig.~\ref{fig4}(d) and (e) show the fit, which is in excellent agreement with the experimental results.


\subsection{Light-activated dynamics}
Finally, we demonstrate that the robot can be programmed to switch between the various dynamics discussed above using light intensity as an external parameter. Since our robot is equipped with two LI sensors at its front and back (Fig. 1(a) $\&$ (b), it continuously measures the value of ambient light intensity, which is set using the overhead projector. We load multiple programs onto the robot's microcontroller and assign light intensity values for each program. For example, we can program the robot to display ABP dynamics when the measured light intensity is between $I_1 - I_2$, BP between $I_3-I_4$, etc., where $I_1$, $I_2$, $I_3$ $\&$, $I_4$ are arbitrary. We summarize this result in SI movie S5, where
a robot is programmed to display BP, ABP, and RTP motion as a function of increasing light intensity.

\section{CONCLUSION}
Here, we introduced a novel experimental system consisting of self-propelled robots. The robots are smart computing devices capable of executing a preloaded program and simulating it to in-plane dynamics with excellent control. Using hardware components like IR and LI sensors, they can also detect obstacles and ambient light intensity and modify their motion as required. The robots are driven by two wheels placed diametrically opposite to each other and move independently with velocities $v_l$ and $v_r$, respectively. We first show that instantaneous values of $v_l$ and $v_r$ calculated from scalar active particle models like Active Brownian, Run and Tumble, and Brownian motion reproduce desired stochastic dynamics in the centre of mass of the robots quite accurately. The experimental parameter space over which this motion can be tuned is quite exhaustive and highly tunable. We also demonstrated the control over robot motion using ambient light as an external control parameter. 

Overall, the experimental system presented in this paper holds great potential in unraveling laws governing the non-equilibrium physics of active and living systems with applications to smart and adaptive material design. In a nutshell, our robots are a simple assembly of electronic and mechanical elements, and we coupled the well-developed theoretical ideas of scalar active particles with the mechanics of such a differential-driven robot. To the best of our knowledge, this is the first-ever study in this direction and carries vast importance considering the rise of the use of programmable particles in active and living matter experiments. The next step is to use a collection of such robots, which offers a significant advantage over already existing experimental systems due to our ability to control individual dynamics with great control. Moreover, we can also engender complex inter-robot interaction using IR and LI sensors which can test not only existing theories in the field but also generate novel dynamical phases hitherto unknown. Furthermore, a collection of such robots also offers possible applications for bio-inspired and nature-inspired robotics.

\section*{ACKNOWLEDGEMENTS}
NK acknowledges financial support from DST-SERB for CRG grant number CRG/2020/002925 and IITB for the seed grant. NK also thanks Sriram Ramaswamy and Sanjib Shabhapandit for productive discussions. AP acknowledges research support from the DST-SERB Start-up Research Grant Number SRG/2022/000080 and the Department of Atomic Energy, India. HS acknowledges SERB for the SRG (grant no. SRG/2022/000061-G). SP thanks CSIR India for research fellowship. \\

\subsection*{APPENDIX A: Image-analysis technique}
In experiments on modeling the robot dynamics, we cover the robot's components (except the light intensity sensors) with a white cap with two black circles drawn on top of it, one bigger than the other, along the robot's orientation. We then use MATLAB's blob analysis to extract their position coordinates. The vector connecting them provides the instantaneous orientation of the robot, and the midpoint of the line connecting them provides the center-of-mass coordinate. The frame rate at which movies are captured varies between 10-100 depending on the robot's speed. The movies are captured for a duration of 1 hour for each model.

\subsection*{APPENDIX B: Removing confinement effect}
Theoretical models of ABP, RTP, and BP dynamics consider an unbounded motion of the particle. In our experiments, we use the following protocol to remove confinement effects caused by boundary walls. Once the robot detects the boundary through its IR sensors, it moves 40 cm in the reverse direction while maintaining its orientation. Later, we remove these straight trajectories from the experiments and stitch all the stochastic tracks together by shifting the coordinates appropriately (See Fig.2 in SI). This creates a continuous free particle motion that is effectively unconfined. \par

\subsection*{APPENDIX C: Theoretical calculation}
\subsubsection{ABP model without translational noise}~\label{theoabp}
In general, the equations of motion for the robot are given by
\begin{subequations}\label{eq1}
\begin{eqnarray}
\dot x(t)&=& v(t) \cos \theta(t),\\
\dot y(t)&= &v(t) \sin \theta(t), \\\label{thetaeq}
\dot \theta(t)&=& \zeta(t),
\end{eqnarray}
\end{subequations}
where $v(t)$ is the component of the instantaneous robot velocity along its orientation $\mathbf{n}=\textbf{(}\cos \theta(t),\sin \theta(t)\textbf{)}$. The values of $v$ and $\zeta$ are varied discretely with the step size of $\epsilon$ second. For mimicking the ABP model, $v$ is kept constant and $\zeta(t)$ is chosen to be a random number with a uniform distribution between $[-\eta,\eta]$. 
For the time scales much larger than $\epsilon$, 
\begin{equation}\label{zetacor}
\left\langle \zeta(t)\zeta(t')\right\rangle =2 D_r \delta(t-t'),
\end{equation}
where $D_r$ is the rotational diffusion constant.
It is straightforward to show that~\cite{balakrishnan2008elements}
\begin{equation}\label{Dreq}
\left\langle \left[ \theta(\tau)-\theta(0)\right]^2 \right\rangle = 2 D_r \tau.
\end{equation}
Here the angular bracket represents the ensemble average. 
Whereas, from the discrete form of~\eqref{thetaeq},
\begin{equation}\label{tau0eq}
\theta(\tau)=\theta(0)+\epsilon\sum^{N_\tau}_{i=1} \zeta_i,
\end{equation}
where $\zeta_i\equiv \zeta( i \epsilon)$ and $N_\tau=\tau/\epsilon$. As the random number $\zeta_i$ has the uniform probability distribution between  $[-\eta,\eta]$,
\begin{equation}\label{key}
\left\langle  \zeta_i\zeta_j\right\rangle =\delta_{ij}\dfrac{\eta^2}{3}.
\end{equation}
Then, from~\eqref{tau0eq},
\begin{eqnarray}\label{key}
\nonumber
\left\langle \left[ \theta(\tau)-\theta(0)\right]^2 \right\rangle&=&\epsilon^2\sum^{N_\tau}_{i=1} \sum^{N_\tau}_{j=1}\left\langle  \zeta_i\zeta_j\right\rangle \\
\nonumber
&=&\epsilon^2 \sum^{N_\tau}_{i=1} \left\langle  \zeta_i^2\right\rangle \\
\nonumber
&=&\epsilon^2N_\tau\dfrac{\eta^2}{3}\\
&=&\epsilon \tau\dfrac{\eta^2}{3}
\end{eqnarray}
Comparing the above equation with~\eqref{Dreq}, we obtain
\begin{equation}\label{key}
D_r=\epsilon \dfrac{\eta^2}{6}.
\end{equation}

\subsubsection{ABP with translational noise}~\label{theoabpNoise}
In this case, $v(t)$ in~\eqref{eq1} is not constant, it rather fluctuates around its mean value. The values of $v(t)$ and $\zeta(t)$ in terms of the velocities of the left and right wheels of the robot, $v_l$ and $v_r$, are given by 
\begin{subequations}\label{vzeta}
\begin{eqnarray}
v(t)&=&\dfrac{v_r(t)+v_l(t)}{2},\\
\zeta(t)&=&\dfrac{v_r(t)-v_l(t)}{d},
\end{eqnarray}
\end{subequations}
where $d$ is the diameter of the robot.
Here $v_l$ and $v_r$ are statistically independent random numbers with a uniform probability distribution in the range   $ [v_a-V,v_a+V]$. Then, one can readily show that 
\begin{eqnarray}\label{key}
\left\langle v_r(i\epsilon) v_r(j\epsilon)\right\rangle &=&\delta_{ij}\dfrac{V^2}{3}+v^2_a,\\
\left\langle v_l(i\epsilon) v_l(i\epsilon)\right\rangle &=&\delta_{ij}\dfrac{V^2}{3}+v^2_a,\\
\left\langle v_r(i\epsilon) v_l(j\epsilon)\right\rangle &=&0.
\end{eqnarray}
Then the relations~(\ref{vzeta}) give
\begin{eqnarray}\label{key}
\label{vivj}
\left\langle v(i\epsilon) v(j\epsilon)\right\rangle &=&\delta_{ij}\dfrac{V^2}{6}+v^2_a,\\
\left\langle \zeta(i\epsilon) \zeta(j\epsilon)\right\rangle &=&\delta_{ij}\dfrac{2V^2}{3d^2},\\\label{vzetacor}
\left\langle v(i\epsilon) \zeta(j\epsilon)\right\rangle &=&0.
\end{eqnarray}
For the time scales much larger than $\epsilon$, it is obvious from~\eqref{vivj} that
\begin{equation}\label{vcor}
\left\langle v(t)v(t')\right\rangle =4 D_t \delta(t-t')+ v^2_a,
\end{equation}
and $\zeta$ follows the correlation relation given by~\eqref{zetacor}. Following the similar steps as done in Sec \ref{theoabp}, one can show that
\begin{eqnarray}\label{key}
D_r&=&\dfrac{\epsilon V^2}{3 d^2},\\
D_t&=&\dfrac{\epsilon V^2}{24}.
\end{eqnarray}
The mean square displacement can then be calculated as follows
\begin{eqnarray}\label{Dreq1}
\nonumber
\Delta r^2(\tau)&=&\left\langle \left[ \mathbf{r}(\tau)-\mathbf{r}(0)\right] \cdot \left[ \mathbf{r}(\tau)-\mathbf{r}(0)\right] \right\rangle\\
\nonumber
&=&\left\langle \int^\tau_0dt\int^\tau_0dt' v(t)v(t') \cos \left[ \theta(t)-\theta(t')\right] \right\rangle \\
&=&\int^\tau_0dt\int^\tau_0dt' \left\langle v(t)v(t') \cos \left[ \theta(t)-\theta(t')\right] \right\rangle  .
\end{eqnarray}
As $v$ and $\zeta$ are statistically independent of each other [from~\eqref{vzetacor}], we have
\begin{eqnarray}\label{Dreq1}
\nonumber
\Delta r^2(\tau)&=&\int^\tau_0dt\int^\tau_0dt' \left\langle v(t)v(t')\right\rangle \left\langle  \cos \left[ \theta(t)-\theta(t')\right] \right\rangle. 
\end{eqnarray}
Using~\eqref{vcor} and the following relation~\cite{vbal1993,santra2020run,howse2007self,basu2018active}
\begin{equation}\label{key}
\left\langle  \cos \left[ \theta(t)-\theta(t')\right] \right\rangle=\exp\left(-D_r \left|t-t' \right| \right),
\end{equation}
we arrive at
\begin{eqnarray}\label{Dreq2}
\nonumber
\Delta r^2(\tau)&=&\int^\tau_0dt\int^\tau_0dt' \Big\lbrace \left[ 4 D_t \delta(t-t')+ v^2_a\right] \\
\nonumber
&&\qquad\times  \exp\left(-D_r \left|t-t' \right| \right)\Big\rbrace\\
&=&4D_t \tau+\dfrac{2 v^2_a}{D^2_r}\left[ D_r\tau+\left(e^{-D_r\tau}-1 \right) \right],
\end{eqnarray}
which was used in Eq. (\ref{eqn5}). To imitate the BP model, we set $v_a=0$ to obtain
\begin{eqnarray}\label{Dreq3}
\Delta r^2(\tau)=4D_t \tau.
\end{eqnarray}
Note that here the dynamics of the robot does not depend on the rotational diffusion constant $D_r$.


\pagebreak
\bibliography{references}

\end{document}